# Magnetic domain scanning imaging using phase-sensitive THz-pulse detection

Finn-Frederik Stiewe[1], Tristan Winkel[1], Tobias Kleinke[1], Tobias Tubandt[1], Hauke Heyen[1], Lucas Vollroth[1], Ulrike Martens[1], Cai Müller[2], Jeffrey McCord[2], Jakob Walowski[1*] and Markus Münzenberg[1]

1 Institut für Physik, Universität Greifswald, Greifswald, Germany
2 Christian-Albrechts-Universität, Kiel, Germany

*Abstract.* In our study, we determine the alignment of magnetic domains in a CoFeB layer using THz radiation. We generate THz-pulses by fs-laser-pulses in magnetized CoFeB/Pt heterostructures, based on spin currents. An LT-GaAs Auston switch detects the radiation phase-sensitively and allows to determine the magnetization alignment. Our scanning technique with motorized stages with step sizes in the sub-micrometer range, allows to image two dimensional magnetic structures. Theoretically the resolution is restricted to half of the wavelength if focusing optics in the far-field limit are used. By applying near-field imaging, the spatial resolution is enhanced to the single digit micrometer range. For this purpose, spintronic emitters in diverse geometric shapes, e.g. circles, triangles, squares, and sizes are prepared to observe the formation of magnetization patterns. The alignment of the emitted THz radiation can be influenced by applying unidirectional external magnetic fields. We demonstrate how magnetic domains with opposite alignment and different shapes divided by domain walls are created by demagnetizing the patterns using minor loops and imaged using phase sensitive THz radiation detection. For analysis, the data is compared to Kerr microscope images. The possibility to combine this method with THz range spectroscopic information of magnetic texture or antiferromagnets in direct vicinity to the spintronic emitter, makes this detection method interesting for much wider applications probing THz excitation in spin systems with high resolution beyond the Abbe diffraction limit, limited solely by the laser excitation area.

Magnetism has become an indispensable part of today's life. In addition to everyday applications in electrical engineering (motors, inductive charging) and medical technology (MRT), the storage of digital data (HDD, racetrack memory) is also based on magnetic phenomena [1–3]. To store data, a surface is permanently magnetized, while to read out the information, the magnetic state of the respective surface is determined by a sensor [3]. Magnetic shape anisotropy also plays an important role in spin alignment [4, 5]. The interaction of the magnetic energy, the exchange- and dipole-dipole interaction leads to the formation of magnetic domains. The alignment of those domains is always configured to minimize the energy in the system. Therefore, domains next to each other may different magnetic orientations. The spatial and temporal arrangement of magnetic domains and their dynamics is in the focus of current research and in demand for new imaging methods [6, 7].
Since the first spin dynamics experiments in 1996 [8, 9], many investigations into the dynamics of magnets have been carried out [10–13]. Those investigations reveal, that small excitations of electrons in thin magnetic films using ultrashort laser pulses generate ultrafast spin currents [14]. Extending ferromagnetic materials (FM) to heterostructures in combination with non-magnetic heavy metals (NM) these excitations can be utilized to emit radiation in the THz frequency range [15–17]. Due to the many useful properties, the demand for the development of concepts and methods utilizing this radiation for imaging is high [16–20]. The scanning technique for THz imaging in combination with near field imaging [18] gains more attention and is used in further investigations resolving objects in the sub micrometer size [21]. The THz pulse phase can be controlled by the magnetization of the emitter [17]. This makes it possible to display structures depending on their magnetization direction and investigate the influence of external magnetic fields on different geometric shapes.

---

* Author to whom correspondence should be addressed: jakob.walowski@uni-greifswald.de

In this study, we use phase-sensitive detection of THz radiation to determine the magnetic state of a FM/NM heterostructure. In order to generate magnetic domains in the ferromagnetic layers of the patterned samples, the strength of the external magnetic field is systematically applied using minor loops. By using a fs laser pulse, we can generate a spatially confined THz-pulse in the heterostructure and thus detect the magnetic alignment in the lithographically patterned THz emitters. The micrometer sized magnetic domains, require the application of near-field optics to image the changes in magnetization direction. In reference [18] we presented the near-field imaging method to overcome the Abbe limit stating, that spatial resolution is limited by half of the involved wavelength, which is larger than 100 micrometers for radiation at 1 THz frequency. We investigate spintronic THz emitters which can be placed in direct proximity to the detector. The strong focusing of the used laser pulses down to the micrometer range confines the spatial expansion of the generated THz source spot to the dimensions of a few micrometers. The test structure consists of many spintronic emitters with different geometric shapes and sizes (from $10\ \mu m$ to $100\ \mu m$), which were lithographically fabricated on a glass substrate. The 2D-scanning technique with motorized stages makes it possible to record a temporal THz spectrum at each position of the sample and thus imaging the patterns. In spite of a smaller achievable spatial resolution compared to magneto optical effects like Kerr- or Faraday-effect, this approach does not rely on large magneto optical response and has the capability for small field change detection in the $\mu T$ range [22].

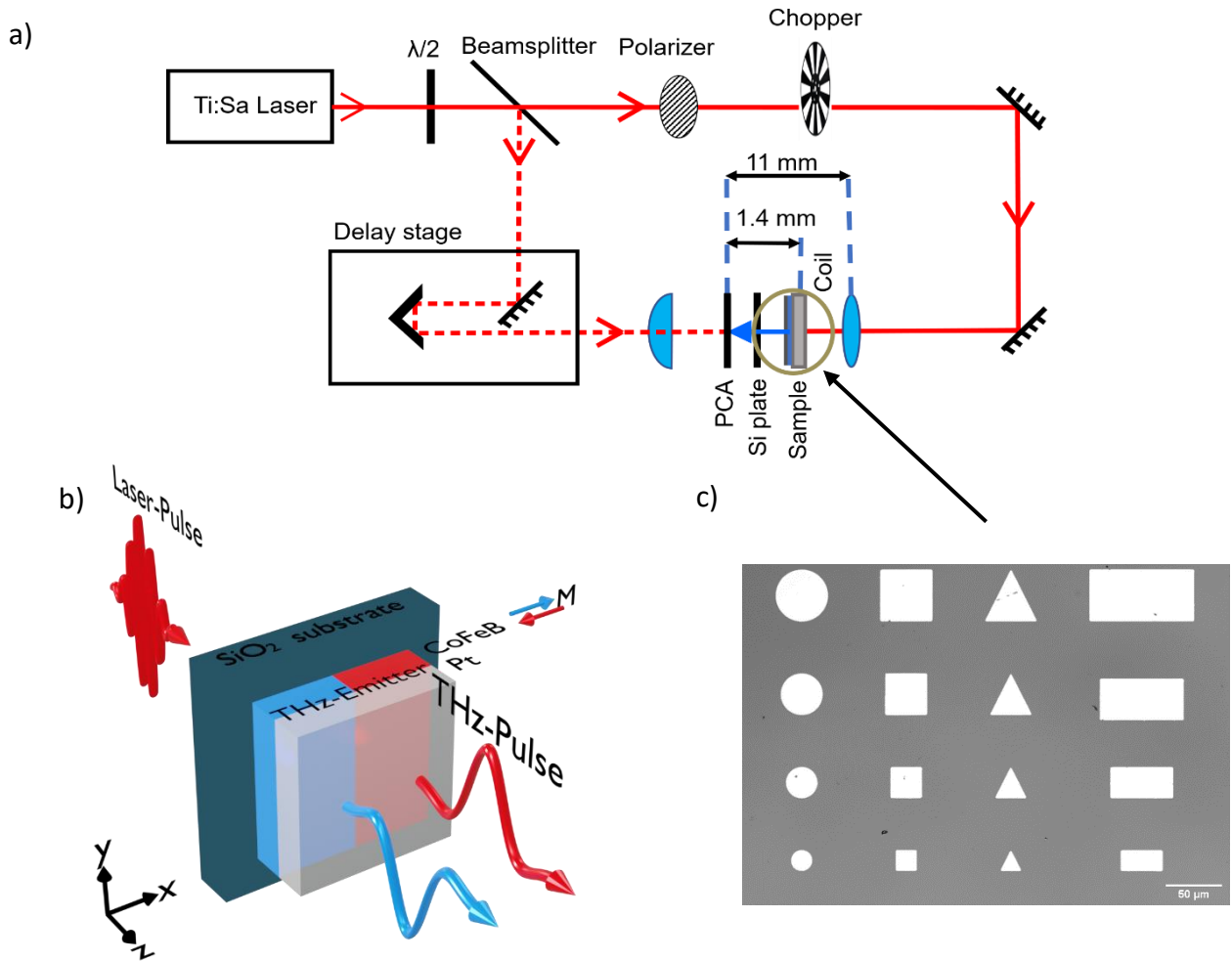

**FIG 1. (a)** Setup: The light source is a Ti:saphire laser system (central wavelength 810 nm, repetition rate 80 MHz, pulse duration 40 fs). Behind a beam splitter, the probe beam is guided to a delay stage to adjust the temporal arrival at the photoconductive antenna detector (PCA) for signal sampling. The pump beam is directed to the spintronic emitter (sample). A polarizer allows laser power adjustment. The signal is recorded via a lock-in amplifier and modulated by a chopper. **(b)** Schematic of a THz-emitter heterostructure. A femtosecond laser pulse excites the ferromagnetic layer (FM) and generates a spin current. Through the inverse spin Hall effect (ISHE), the current is converted into a transient charge current in the non-magnetic layer (NM). The emitted THz has a different phase depending on the magnetization, shown in blue and red. **(c)** Microscopic image showing the investigated geometrical structures on the sample (light grey). They act as spintronic emitters and are the same time as magnetic structures and have sizes ranging from $100\ \mu m$ down to $10\ \mu m$.

The optical setup for detection and imaging magnetic domains is shown schematically in Figure 1 (a). The operation principle is explained in detail in [18]. For THz-radiation generation, a Ti:Sa laser (Coherent Vitara, 40 fs pulse duration, 810 nm central wavelength, 80 MHz repetition rate) is used. A beam splitter divides the laser into a lower-powered probe (dashed line) and a higher-powered pump beam (solid line). The probe beam is guided to an Auston switch (photoconductive antenna, PCA) through a delay stage. This PCA acts as a THz-detector. The laser pulse short-circuits the antenna by exciting electrons in the PCA gap. The pump beam is guided to the THz emitter sample. A polarizer in the beam path enables laser power adjustment. A chopper modulates the pump beam with a frequency of 1.5 kHz. The beam is focused onto the sample, which emits THz radiation generated in the CoFeB/Pt heterostructures (details in [14, 18, 23, 24]). Helmholtz coils are placed around the sample to control the magnetization state by applying magnetic fields up to $4\,\text{mT}$ and continuously reducing the field strength. This sample demagnetization using minor loops generates magnetic domains. They are imaged by scanning the sample with two motorized μm-stages moving the heterostructure horizontally and vertically (x- and y-direction) against the pump-beam with minimal step sizes of $200\,\text{nm}$. The emitted THz pulses are sampled by delaying the probe beam in time and the amplified signal is recorded by a lock-in amplifier.

The data is analyzed using the following procedure. A complex value $z$ consisting of $x + yi$ from the data are recorded by the lock-in amplifier. In the first step, we calculate the phase of the real signal in the complex $z$ values. This is done by transforming the $z$ values into polar coordinates, performing a discrete Fourier transform, and picking the phase of the value representing the second harmonic. The original $z$ values are rotated by subtracting the obtained phase, setting the imaginary part to 0 and leaving only the real part of $z$ for further calculation. In the second step, the signal is transformed using Fast Fourier Transform (FFT) to obtain the frequency spectra contained in the THz signals. The resulting THz frequency spectra have again a real and a complex component. By determining the phase and rotating it, the signal, shifted to the real component. In general, we use the $1\,\text{THz}$ amplitude from each recorded THz spectrum combined into one image to represent the magnetization state.

The investigated sample is schematically shown in figure 1(b). Figure 1(c) shows a microscope image of the spintronic emitters (bright forms) patterned into different shapes and sizes from a film, deposited on a glass substrate. To create a spintronic emitter, a $2\,\text{nm}$ thick CoFeB layer is magnetron sputtered onto the substrate, directly followed by a $2\,\text{nm}$ Pt layer deposited by e-beam evaporation without breaking the vacuum conditions. After thin film deposition, the layer stack is coated with a negative resist (AR-N4340) and subsequently baked on a hot plate. The resist was exposed using a laboratory mask aligner (MJB4, Süss MicroTec) with a wavelength of $405\,\text{nm}$. Afterwards, the unexposed resist is removed by development with AR 300-47. The uncovered, resist-free areas are Ar-etched in a home-built system. This leaves spintronic emitters with various geometrical shapes on the glass substrate as shown in figure 1(c).

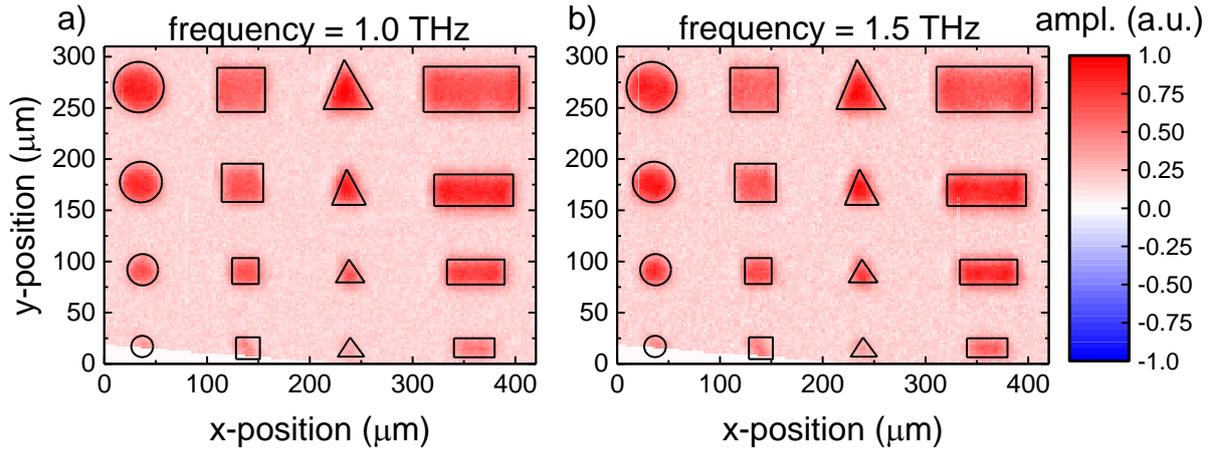

**FIG 2.** Two-dimensional THz images of the magnetized sample on an $450\ \mu m \times 300\ \mu m^2$ area scanned with a step size of $s_{xy} = 2\ \mu m$ extracted for the frequencies **(a)** $f = 1.0\ THz$ and **(b)** $f = 1.5\ THz$. The THz emitting structures can be seen clearly for all included sizes. The images are rotated by $4.8°$, to align the structures for better visibility.

Figure 2 depicts two-dimensional THz images of the test pattern (figure 1 (c)) for two extracted frequencies $f = 1.0$ THz and $f = 1.5$ THz. The images are obtained by scanning a sample area of $420 \times 310$ μm in x- and y-direction, with a step size of $s_{xy} = 2$ μm. The white area in the lower left corner originates from a rotation of the image. During the measurement, the sample was slightly canted by 5°, and rotated afterwards for a better visibility. A few data points are missing in the lower left corner indicated by the white area.

At each position, a temporal spectrum of the THz pulse is recorded. By using Fast Fourier Transformation (FFT), the included frequencies are extracted from each spectrum, as described above. We show normalized THz e-field amplitudes $A$ for the frequencies $f = 1.0$ THz (figure 2 (a)) and $f = 1.5$ THz (figure 2 (b)). The geometrical structures are clearly imaged in red color and indicated by the black edges for a better visibility. The environment around these structures, where the CoFeB/Pt layer is removed, only show a small amplitude within the noise level (light red). In general, the larger shapes, show a larger $A$ than the smaller ones. This is a result of the small structure size compared to the scanning laser beam. For both frequencies the circular structures exhibit a slightly larger THz signal amplitude than the other geometrical shapes, because the shape anisotropy is not direction dependent and the sample was lightly canted during measurement while applying only a small field of $4$ mT.

Figure 3 shows an overview scan of an $450 \times 350$ μm area obtained with a step size of $s_{xy} = 10$ μm. The magnetic structures are imaged after demagnetization using minor loops to create magnetic domains for two distinct frequencies $f = 1.0$ THz and $f = 1.5$ THz. The Helmholtz coils are placed around the sample, as indicated in the schematic in figure 1(a) to align the applied field direction along the y-position axis figures 2 and 3 and in the plane of the magnetic sample.

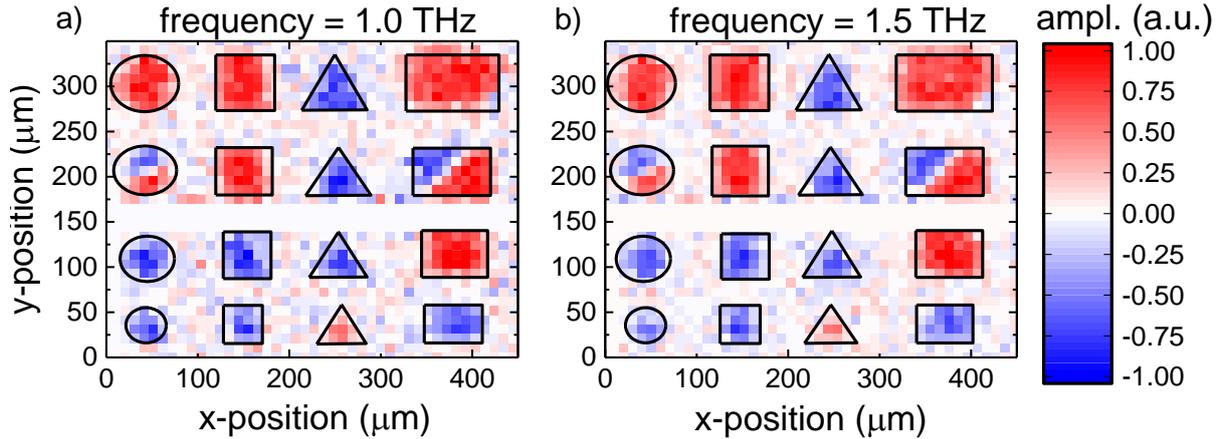

**FIG 3.** *Two-dimensional THz images with a step size of $s_{xy} = 10\ \mu m$ of the structures after the minor loop demagnetization process for the frequencies **(a)** $f = 1.0\ THz$ and **(b)** $f = 1.5\ THz$. Two structures, one circle and one rectangle show domain formation. The other structures switched their magnetization, randomly, as the applied field step resolution was not sufficient to halt the switching at pinning centers.*

Our phase sensitive detection method shows, the formation of domains with opposite magnetization in several geometrical structures in figure 3(a) and (b). The overview scan was carried out in two parts. The data points around y-position $150\ \mu m$ are missing, because the second part was started at a lower position, leaving out the blank area without magnetic structures.

Creating magnetic domains in the structures is very sensitive to the applied field resolution and settings. Obtaining a resolution around $0.1\ mT$ is not sufficient to reliably create identic magnetic domains after each procedure. Some structures magnetizations switched entirely applying those field steps in one or the other direction. This shows in the apparent random magnetization orientation for the involved structures. By applying a positive magnetic field, all patterns show positive THz e-field amplitudes (see figure 2). Reversing the magnetic field flips the spins and switches to a negative THz E-field amplitude. However, two structures exhibit enough pinning centers inside, to create domain walls within the applied field parameters. This is shown in figures 3 (a) and 3 (b) in the second row, for the circle and the rectangle structure. In this data, the domain walls can be seen as white spaces, with a detected THz e-field amplitude averaged to zero, because within this course resolution, the THz pulse is generated within an area, where the magnetization is pointing in opposite directions. There, the generated THz pulse has two components shifted by 180° resulting in a zero signal at the detector. Besides this, the other structures show decreased THz e-field amplitudes, especially on the edges. That hints to a partial domain wall formation, which cannot be resolved in this coarse depiction.

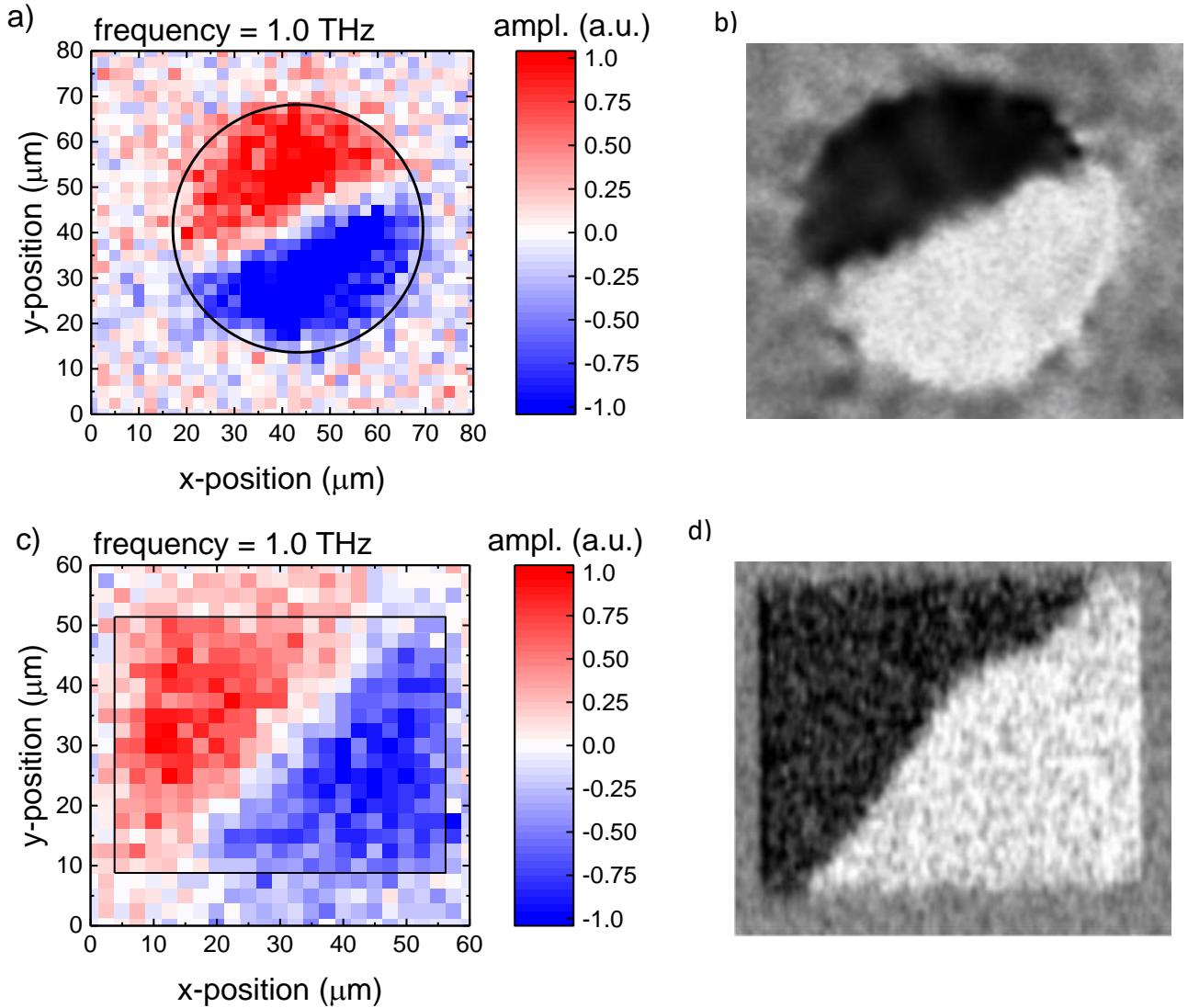

***FIG 4.*** *Detailed THz imagines of the circle **(a)** and rectangle **(c)** after domain wall formation extracted for $f = 1.0\ THz$. For comparison, both structures, the circle and the rectangle are shown imaged by Kerr microscopy in **(b)** and **(d)** respectively.*

Figure 4 shows the two structures exhibiting magnetic domains scanned in more detail using a smaller step size of $s_{xy} = 2.5$ µm. The 2D THz scans of the circle (figure 4 (a)) and the rectangle (figure 4 (c)) show the bisecting of the geometric shapes by a domain wall (white). The detected THz signal adds up to zero in the domain wall region, because of the magnetization rotation in this area. The structures are divided in a positive (red) and a negative (blue) domain magnetization pointing in opposite directions. The THz e-field amplitude in the domain magnetized in negative direction is slightly larger than in the positive direction. In general, the THz e-field Amplitudes are smaller for the demagnetized samples, due to the smaller magnetization in the remanent state and thus a smaller electron polarization involved in the process generating the inverse spin-Hall effect. In addition, changing the magnetization back and forth, by applying minor loops blurs the remaining magnetization in the remanent state leading to different THz e-field amplitudes arriving at the detector for each magnetization direction. Furthermore, figures 4 (b) and (d) show Kerr images of the corresponding structures shown in figures 4 (a) and (c) to ensure the THz images show oppositely magnetized domains instead of a phase shift arising from the instability of other experimental parameters during the measurement. Those could be e.g. thermal expansion or instabilities leading to a shift of the laser beam. However, both imaging techniques show the same magnetic alignment.

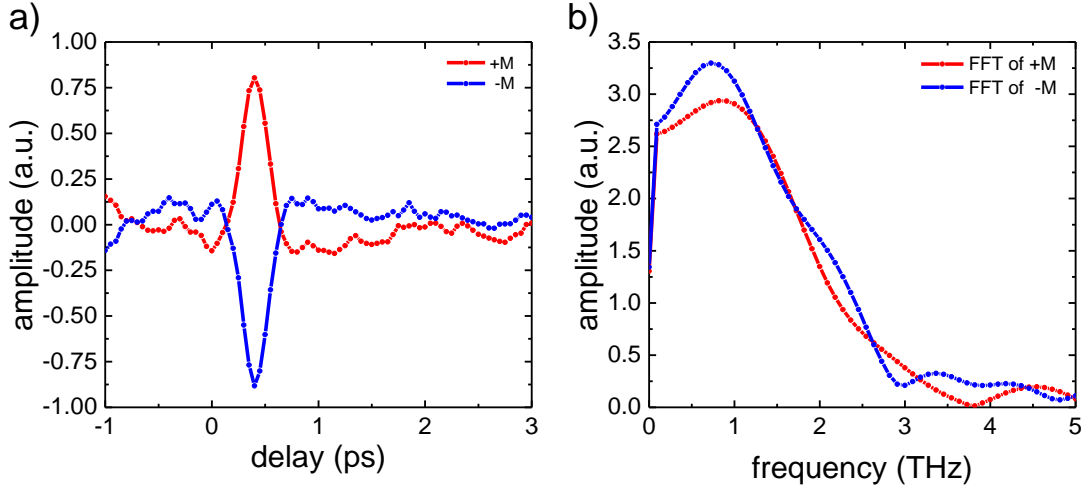

*FIG 5. (a) THz signal recorded in a magnetic field **M**. Both signals recorded by saturating the magnetization in opposite magnetic fields of 4 mT clearly show a 180° phase shift. (b) The Fourier Transformed THz frequency amplitudes after zeropadding show a slight difference due to signal fluctuations.*

Figure 5 shows the phase shift in the THz e-field amplitude arising from the magnetization direction for two opposite magnetizations +**M** (red) and -**M** (blue). The THz e-field peak is detected at the same delay position but with opposite amplitude sign corresponding to a phase shift by 180°. The signal is recorded by saturating the magnetic structure with an applied field in both magnetization directions. The corresponding FFT for +**M** and -**M** are depicted in figure 5(b). Both magnitudes of the THz e-field amplitude are approximately the same, because the magnetization vector is parallel to the easy plane of the FM thin film. The difference stems from fluctuations during the measurement.

The behavior of the THz-amplitudes in the vicinity of a domain wall is shown in more detail in figure 6 (a) for the circle structure. The measured THz spectra at seven positions across a domain wall at a distance of 15 µm. The data shows, that at the domain wall center (position 4), the THz e-field amplitude is zero for the whole spectrum. At this position, the laser pulse excites electrons within the domain wall and due to its size in both domains. Therefore, two spin currents with opposite spin polarization are generated and the resulting THz amplitude adds up to zero. Moving away from the domain wall center by 7.5 µm, towards position 1 and 7, the amplitude increases and reaches its maximum at those positions. This corresponds to full spin polarizations reached when a single magnetization area is involved in the THz generation process. For both magnetization directions shown in red and blue, the THz e-field spectra are phase shifted, as expected for opposite spin polarizations. The data clearly shows that magnetization change can be detected, and magnetic domains can be allocated, including the domain wall positions.

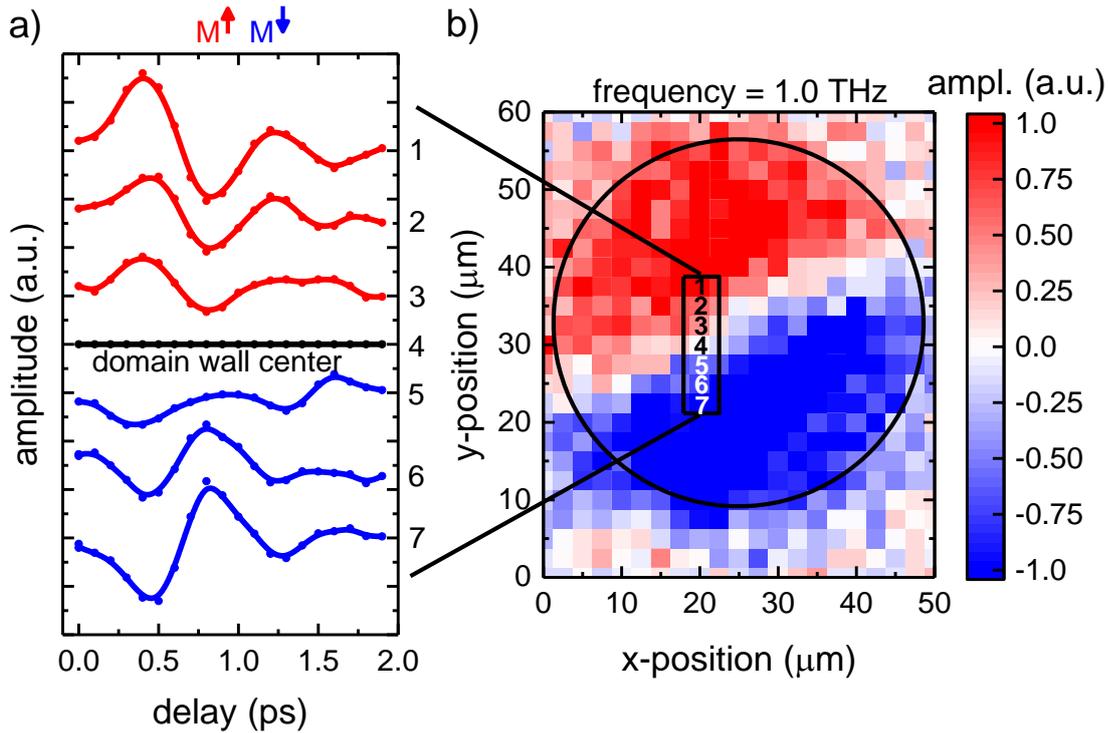

*FIG 6. (a) THz signals at different positions crossing a domain wall. The cross section is 15 μm divided into seven steps with the domain wall center at position 4. The red graphs are magnetized up (positive), the blue ones are oppositely magnetized down (negative). At (1) and (7) the peak-to-peak value of the THz amplitude reaches its maximum, because there, the magnetization is homogeneous in one direction. At smaller distance from the domain wall, the THz e-field amplitude decreases, with a changing magnetization. The extracted THz amplitudes for $f = 1.0$ THz are marked in the THz image (b).*

The results of this study show one application for phase-sensitive THz pulse detection using spintronic THz emitters. The emitters can be deposited and structured directly on a glass substrate and can therefore be implemented easily and in a space-saving manner. This ensures identical conditions for THz generation at each emitter-spot and makes it possible to image arbitrary shapes in 2-dimensions, as far as they are lithographically available. The possibility to generate spin currents in differently magnetized domains simultaneously allows to emit THz radiation with phase shifted e-field amplitudes. In combination with a rotation of the sample, this method can be extended to image and the domain wall region and investigate the magnetization state. Further, magnetic structures with a large shape anisotropy, like stripes arrays can be structured in a 90° angle to each other. This would allow to emit THz waves with e-field amplitudes oscillating perpendicularly. The only missing component would be an adjustable temporal delay between both perpendicular waves to create phase-controlled THz radiation like when using a retardation plate. This would allow to generate circular polarized THz radiation of any shape. Those possibilities make this imaging technique a new versatile tool for magnetic spectro-microscopy in the THz region with high spatial resolution. Imaging magnetization of very small areas fast and without affecting the alignment. A further potential is in the imaging of magnetic insulators which could be attached in close vicinity to the emitter, using the generated THz radiation to determine their magnetization state in transmission.

In the future it will be possible to study the magnetization of different shapes depending on their thickness and their influence on the THz radiation. Implementing special collimation lenses, will enable to examine even smaller areas and improve the spatial resolution. In addition, by changing the detector to a polarization-independent THz detector, a more detailed representation of the domain walls and the areas in which the rotation of the magnetization takes place wil be realized. The useable frequency range can be extended by implementing new techniques for detecting THz radiation, for example newer generations of PCA or electro-optical detectors such as ZnTe crystals. This makes it possible to study the influence of magnetization in the full frequency range, including higher frequencies ($> 3\,\text{THz}$).

The authors gratefully acknowledge the financial support from the BMBF, MetaZIK PlasMark-T (No. FKZ:03Z22C511).

## AUTHOR DECLARATIONS
**Conflict of Interest**
The authors have no conflicts of interest to disclose.

## DATA AVAILABILITY
The data that support the findings of this study are available from the corresponding author upon reasonable request.

# References

1. S. Parkin, S.-H. Yang, Nature nanotechnology **10**, 195 (2015)
2. S.S.P. Parkin, M. Hayashi, L. Thomas, Science (New York, N.Y.) **320**, 190 (2008)
3. R. Wood, Journal of Magnetism and Magnetic Materials **321**, 555 (2009)
4. Dubowik, Physical review. B, Condensed matter **54**, 1088 (1996)
5. M. Grimsditch, Y. Jaccard, I.K. Schuller, Physical review. B, Condensed matter **58**, 11539 (1998)
6. T. Eggebrecht, *Lichtinduzierte magnetische Defekte in ultradünnen Filmen* (2018)
7. T. Eggebrecht, M. Möller, J.G. Gatzmann, N. Da Rubiano Silva, A. Feist, U. Martens, H. Ulrichs, M. Münzenberg, C. Ropers, S. Schäfer, Physical review letters **118**, 97203 (2017)
8. Beaurepaire, Merle, Daunois, Bigot, Physical review letters **76**, 4250 (1996)
9. A. Scholl, L. Baumgarten, R. Jacquemin, W. Eberhardt, Phys. Rev. Lett. **79**, 5146 (1997)
10. R. Zhou, Z. Jin, G. Li, G. Ma, Z. Cheng, X. Wang, Appl. Phys. Lett. **100**, 61102 (2012)
11. A.V. Kimel, A. Kirilyuk, A. Tsvetkov, R.V. Pisarev, T. Rasing, Nature **429**, 850 (2004)
12. C.D. Stanciu, F. Hansteen, A.V. Kimel, A. Kirilyuk, A. Tsukamoto, A. Itoh, T. Rasing, Physical review letters **99**, 47601 (2007)
13. T.J. Yen, W.J. Padilla, N. Fang, D.C. Vier, D.R. Smith, J.B. Pendry, D.N. Basov, X. Zhang, Science (New York, N.Y.) **303**, 1494 (2004)
14. M. Battiato, K. Carva, P.M. Oppeneer, Physical review letters **105**, 27203 (2010)
15. T. Seifert, U. Martens, S. Günther, M.A.W. Schoen, F. Radu, X.Z. Chen, I. Lucas, R. Ramos, M.H. Aguirre, P.A. Algarabel, A. Anadón, H.S. Körner, J. Walowski, C. Back, M.R. Ibarra, L. Morellón, E. Saitoh, M. Wolf, C. Song, K. Uchida, M. Münzenberg, I. Radu, T. Kampfrath, SPIN **07**, 1740010 (2017)
16. J. Walowski, M. Münzenberg, Journal of Applied Physics **120**, 140901 (2016)
17. T.S. Seifert, L. Cheng, Z. Wei, T. Kampfrath, J. Qi, Appl. Phys. Lett. **120**, 180401 (2022)
18. F.-F. Stiewe, T. Winkel, Y. Sasaki, T. Tubandt, T. Kleinke, C. Denker, U. Martens, N. Meyer, T.S. Parvini, S. Mizukami, J. Walowski, M. Münzenberg, Appl. Phys. Lett. **120**, 32406 (2022)
19. E.T. Papaioannou, R. Beigang, Nanophotonics **10**, 1243 (2021)
20. P. Klarskov, H. Kim, V.L. Colvin, D.M. Mittleman, ACS Photonics **4**, 2676 (2017)
21. P. Li, S. Liu, Z. Liu, M. Li, H. Xu, Y. Xu, H. Zeng, X. Wu, Appl. Phys. Lett. **120**, 201102 (2022)
22. D.S. Bulgarevich, Y. Akamine, M. Talara, V. Mag-usara, H. Kitahara, H. Kato, M. Shiihara, M. Tani, M. Watanabe, Sci Rep **10**, 1158 (2020)
23. E. Saitoh, M. Ueda, H. Miyajima, G. Tatara, Appl. Phys. Lett. **88**, 182509 (2006)
24. T. Seifert, S. Jaiswal, U. Martens, J. Hannegan, L. Braun, P. Maldonado, F. Freimuth, A. Kronenberg, J. Henrizi, I. Radu, E. Beaurepaire, Y. Mokrousov, P.M. Oppeneer, M. Jourdan, G. Jakob, D. Turchinovich, L.M. Hayden, M. Wolf, M. Münzenberg, M. Kläui, T. Kampfrath, Nature Photon **10**, 483 (2016)